\documentclass[12pt]{article}
\def\spose#1{\hbox to 0pt{#1\hss}}
\def\simlt{\mathrel{\spose{\lower 3pt\hbox{$\mathchar"218$}}
     \raise 2.0pt\hbox{$\mathchar"13C$}}}
\def\simgt{\mathrel{\spose{\lower 3pt\hbox{$\mathchar"218$}}
     \raise 2.0pt\hbox{$\mathchar"13E$}}}

\usepackage{psfig}
\usepackage{epsfig}
\usepackage{amssymb}
\begin{document}
\title{Deriving the Metallicity Distribution Function of Galactic Systems} 
\author{Yeshe Fenner $^{1}$ \&
Brad K. Gibson  $^{2}$
} 
\date{}
\maketitle
{\center
$^1$ Centre for Astrophysics \& Supercomputing,
Swinburne University, Mail \#31, Victoria, Australia, 3122\\yfenner@astro.swin.edu.au\\[3mm]
$^2$ Centre for Astrophysics \& Supercomputing,
Swinburne University, Mail \#31, Victoria, Australia, 3122\\bgibson@astro.swin.edu.au\\[3mm]
}
\begin{abstract}
  
The chemical evolution of the Milky Way is investigated using a
dual-phase metal-enriched infall model in which primordial gas fuels
the earliest epoch of star formation, followed by the ongoing
formation of stars from newly accreted gas. The latest metallicity
distribution of local K-dwarfs is reproduced by this model, which
allows the Galactic thin disk to form from slightly metal-enriched
gas with $\alpha$-element enhancement. Our model predicts ages for
the stellar halo and thin disk of 12.5 and 7.4~Gyr, respectively, in
agreement with empirically determined values. The model presented in
this paper is compared with a similar dual-phase infall model from
Chiappini et al. (2001). We discuss a degeneracy that enables both
models to recover the K-dwarf metallicity distribution while
yielding different star formation histories.
  
The metallicity distribution function (MDF) of K-dwarfs is proposed
to be more directly comparable to chemical evolution model results
than the G-dwarf distribution because lower mass K-dwarfs are less
susceptible to stellar evolutionary effects. The K-dwarf MDF should
consequently be a better probe of star formation history and provide
a stronger constraint to chemical evolution models than the widely
used G-dwarf MDF. The corrections that should be applied to a
G-dwarf MDF are quantified for the case of the outer halo of
NGC~5128.

\end{abstract}


\bigskip

\section{Introduction}

The metallicity distributions of stars in different environments yield
important information about the age and formation history of stellar
systems. Particular attention has been paid to the Metallicity
Distribution Function (MDF) of G-dwarfs in the solar neighbourhood, as
this provides one of the strongest constraints on Galactic Chemical
Evolution models. These models solve the sets of
differential equations governing the formation and destruction of the
elements through stellar processes (Tinsley 1980). The over-prediction
of low-metallicity stars, the so-called ``G-dwarf problem'', is
typically resolved by allowing the disk of the Milky Way to form
gradually from accreting gas on a timescale of about 7~Gyr. Recent
models (e.g. Chiappini et~al. 1997; Goswami \& Prantzos 2000) have
successfully reproduced the observed G-dwarf MDF using dual-phase
accretion models in which the halo/thick disk component evolves
independently of the thin disk and on a rapid timescale.

An implicit assumption of most theoretically predicted G-dwarf
metallicity distributions is that all G-dwarfs have lifetimes older
than the age of the Milky Way. In fact, the stars in the Rocha-Pinto \&
Maciel (1996) sample have masses ranging from 0.7 M$_{\odot}$ to 1.1
M$_{\odot}$, while Wyse \& Gilmore's (1995) dataset spans 0.8
M$_{\odot}$ to 1.2 M$_{\odot}$. Some of these stars have lifetimes
shorter than the age of the disk. Furthermore, the earliest formed
G-dwarfs will have even shorter lifetimes owing to their low
metallicity, which reduces their opacity and raises their luminosity. For
instance, a 1.3 M$_{\odot}$ star leaves the main-sequence after about
4 Gyr at solar metallicity, or after less than 3 Gyr for
Z~=~0.05~Z$_{\odot}$ (Schaller et~al. 1992). In the low-metallicity
environment of the early Milky Way, a 1~M$_{\odot}$ star is expected
to leave the main-sequence after about 6 Gyr. The preferential loss of
the oldest G-dwarfs from empirical datasets tends to bias the MDF
towards higher metallicity and suppresses the low-metallicity tail.
However, the protracted formation history of the Galactic disk means
that stellar lifetime effects alone cannot account for the paucity of
metal-poor G-dwarfs with respect to ``closed box'' models of chemical
evolution. We have investigated the impact of these effects on the
shape and peak of measured MDFs in different stellar environments,
the results of which are presented in Section~\ref{Harris}. 
Rocha-Pinto \& Maciel (1997) and Bazan \& Mathews (1990)
should be referred to for additional commentary in this regard.

Since K-dwarfs are less massive than G-dwarfs, their lifetimes are longer 
than the age of the Galaxy. The K-dwarf metallicity distribution should
therefore trace the ``true'' MDF because surveys of this stellar
population should not suffer from incompleteness due to stars evolving
off the main-sequence. Because of their faintness, however, accurate
metallicities of large numbers of K-dwarfs have only recently been
obtained. Kotoneva et~al. (2002) constructed a K-dwarf
distribution using a sample of 220 nearby K-dwarfs drawn from the
Hipparcos catalogue (ESA 1997). They were able to select K-dwarfs by
absolute magnitude and restrict the mass range to 0.7 - 0.9
M$_{\odot}$, for which evolutionary effects are unimportant.

Galaxy Evolution tool (Fenner \& Gibson 2003) - {\tt
  GEtool}\footnote{\tt http://astronomy.swin.edu.au/GEtool/} - was
used to numerically solve the equations governing the rate of change
of gas mass and elemental abundances due to star formation, supernovae
feedback, and the continual accretion of gas onto the disk. The model
presented here is capable of reproducing the main observed properties
of the Milky Way, including the present day radial gas distribution,
radial abundance gradients, age-metallicity relation, stellar
metallicity distribution and elemental abundance ratios.
Section~\ref{model} outlines the features of the model, while
Section~\ref{yields} describes the adopted stellar nucleosynthetic
prescriptions.  The predicted K-dwarf metallicity distribution is
presented in Section~\ref{results}, where it is compared against the
measured K-dwarf MDF in order to infer the possible formation history
of the Galaxy. In Section~\ref{Harris} are shown results from a study
into the effect of finite stellar lifetimes on the measured MDFs of
populations in different star forming environments.

\section{The Model}\label{model}

\subsection{Infall scheme}
Our Galactic evolution model assumes two main formation phases. The
formation of halo stars is associated with the first episode of
infalling material, while the second phase leads to disk formation. To
aid the comparison with other dual-phase Galactic chemical evolution
studies the model is restricted to simulate \emph{two} rather than
\emph{three} infall phases. The thick disk is assumed to correspond to
a merger heated thin disk and is therefore chiefly associated with the
early stages of the second formation epoch. The difficulty inherent in
using this kind of dual-phase model to describe a three component
system is further discussed in Section~\ref{results}. Simulating the
evolution of a separate thick disk component will be explored in
forthcoming work.

The initial phase of star formation occurs on a timescale of $\sim
0.5$\,Gyr and enriches the initially primordial gas to a metallicity
of [Fe/H]$\approx$$-$1.2. The second formation phase is delayed by
1\,Gyr with respect to the first phase and has a more protracted star
formation history; High-Velocity Clouds (HVCs) may represent the
present-day source of this Galactic star formation fuel.  Drawing on
observations of the chemical composition of these clouds (Wakker
et~al. 1999; Gibson et~al. 2001; Sembach et~al. 2002), the second
accretion phase assumes the gas is slightly metal-enriched
(Z\,=\,0.1\,Z$_\odot$) with an enhancement of $\alpha$-elements
relative to iron. The adopted value of [$\alpha$/Fe]\,=\,$+$0.4 is
consistent with $\alpha$/Fe ratios found in metal-poor stars (e.g.
Ryan, Norris \& Beers, 1996). The precise dependence of key
observational constraints on the assumed chemical composition of the
infalling gas will be presented in a forthcoming paper (Fenner \&
Gibson 2003).

After Chiappini et~al. (1997), we assume that the
rate at which material is accreted during these phases declines
exponentially. The evolution of total surface mass density
$\sigma_{tot}(r,t)$ is given by
\begin{equation}
\frac{d \sigma_{tot}(r,t)}{dt} = A(r) e^{-t/\tau _H(r)} + B(r)
e^{-(t-t_{delay})/ \tau _D(r)}
\end{equation}
\noindent
where the infall rate coefficients $A(r)$ and $B(r)$ are chosen in
order to reproduce the present-day surface mass density of the halo
and disk components, which we take to be 10 and
45\,M$_{\odot}$\,pc$^{-2}$, respectively. These values are comparable
with 17 and 54\,M$_{\odot}$\,pc$^{-2}$ used by Chiappini et al.
(2001). The adopted timescales for the infall phases are
$\tau_H$\,=\,0.5\,Gyr and $\tau_D$\,=\,7.0\,Gyr at the solar radius
$r_{\odot}$\,=\,8\,kpc.  The ``inside-out'' functional form for
$\tau_D(r)$ (Romano et~al. 2000) is adopted, with the Milky Way age
taken to be 13\,Gyr.

\subsection{Star formation formalism}
We adopt a star formation prescription which is based upon the
hypothesis that star formation is
triggered by the compression of interstellar material from spiral arm
motion (e.g. Prantzos \& Silk 1998). This implies that the star
formation rate (SFR) is proportional to the angular frequency of the
spiral pattern and therefore inversely proportional to radius. The
star formation law incorporated in this model is given by
\begin{equation}\label{SFR}
\psi (r,t) =  \nu \, \sigma_{gas}^{2}(r,t) \, \Big(
\frac{r_{\odot}}{r} \Big) \end{equation}
\noindent
where the value of the efficiency factor $\nu$ is constrained by the
present day gas fraction.

\subsection{Initial mass function}
The shape of the stellar initial mass function (IMF) influences the
quantity of Galactic material locked up in stars of different masses,
which in turn determines the rate at which different elements are
released into the interstellar medium.  The models presented in
Section~\ref{results} use the Kroupa et~al. (1993) three-component
IMF.  With respect to the Salpeter (1955) and Scalo (1986) IMFs, this
function has a steeper slope for high masses, resulting in less
material being processed by massive stars.  We impose an upper mass
limit of 60\,M$_\odot$ on stellar formation in order to recover the
observed trend of [O/Fe] at low metallicity (e.g. Carretta et al. 2000, Melendez et al. 2001) while avoiding the
overproduction of oxygen.

\subsection{Properties of the halo, thick disk, and
  thin disk components} 

The ratio of the local number density of thick disk to thin disk stars
lies between 1\% and 6\% (Gilmore et~al.  1995). Both disk components
are well-fitted in vertical height, $z$, with exponential functions. Taking the thick
and thin disk scale heights to be 1~kpc and 330~pc, respectively
(Gilmore et~al. 1995), the ratio of the thick to thin disk surface
densities lies between 2\% and 20\%. The local stellar density of halo
stars relative to thin disk stars is only $\sim 0.1$\% with a mean
metallicity of [Fe/H]\,=\,$-$1.7 (Norris \& Ryan 1991), while thick
disk stars have typical metallicities in the interval $-$0.4 $<$
[Fe/H] $<$ $-$1.2 with a mean of $-$0.7 (Gilmore et~al.  1995). In
comparison, the MDF of thin disk stars peaks near [Fe/H]\,=\,$-$0.2
(McWilliam 1997).  In addition to metallicity, kinematic behaviour is
used to distinguish between these three components. Thick disk
kinematics lie somewhere between those of the halo and the thin disk,
and describe a distinct population.

\section{Yields}\label{yields}

The observed behaviour of elemental ratios such as [O/Fe] with [Fe/H]
are important observational constraints on chemical evolution models.
Oxygen is chiefly produced in massive stars, whereas a significant
fraction of iron is supplied by Type~Ia supernovae involving binary
systems of lower mass stars. The characteristic timescale on which
oxygen is ejected into the ISM is therefore much shorter than that of
iron, owing to the mass dependence of main-sequence lifetimes.
 
In order to recover the observed behaviour of elemental abundance
ratios, the instantaneous recycling approximation has been relaxed.
Mass and metallicity-dependent lifetimes were taken from Schaller et
al. (1992).

\subsection{Massive stars}
The present work incorporates an updated set of stellar yields from
Limongi et~al. (2000; 2002), supplemented with a finer mass
coverage grid (kindly provided by Marco Limongi), covering a range of
stellar initial masses and metallicities (13$<$$m/$M$_{\odot}<$80 and
Z/Z$_\odot$=0, 10$^{-3}$, 1). Stellar yields are one of the most
important ingredients in Galactic chemical evolution models, yet
questions remain concerning the composition of ejected material, owing
to the uncertain role played by processes such as mass loss, rotation,
fall-back, and the location of the mass cut in supernovae, which
separates the remnant from the ejected material.

The amount of iron released in supernova explosions of massive stars
is particularly sensitive to fall-back and the location of the mass
cut.  There are few supernova observations with which to directly
infer the amount of iron ejected. We have therefore taken the opposite
approach of using the observed trend of abundance ratios to indirectly
constrain iron yields.  The Limongi et~al.  (2000; 2002) yields
are presented for a range of mass cuts, allowing a self-consistent
treatment of all yields (c.f.  Goswami \& Prantzos 2000; Argast et~al.
2002).  We have found that the trend of [O/Fe] with [Fe/H] is best
recovered by suppressing the ejection of iron in stars more massive
than about 30~M$_{\odot}$.  This leads to a smaller contribution to
the interstellar iron abundance from Type~II supernovae than is
obtained from the canonical Woosley \& Weaver (1995) model.  A
preliminary relationship between the mass cut and iron yield was
presented in Fenner, Gibson \& Limongi (2002); a full accounting will
be provided in Fenner \& Gibson (2003).

\subsection{Low- and intermediate-mass stars}
The metallicity-dependent yields of Renzini \& Voli (1981) were used
for stars in the 1\,$\le$\,m/M$_\odot$\,$\le$8 range. Yields for stars with
masses between 8~M$_{\odot}$ and 13~M$_{\odot}$ were estimated by
interpolating between the highest mass in Renzini \& Voli and the
lowest mass in Limongi et~al (2000; 2002). 
It should be noted, however, that it is not yet
clear whether stars in the range 8\,$<$\,m/M$_{\odot}$\,$<$\,13 contribute to
interstellar enrichment, or whether the processed material becomes
trapped in the remnant following core-collapse.

\subsection{Type Ia supernovae}
The yields for Type Ia SNe were taken from the W7 model of Thielemann
et~al. (1993). The fraction of binaries resulting in Type~Ia supernovae was
fixed by requiring that the model reproduce the local age-metallicity
relation; a fraction of 3\% was adopted, leading to model predictions
consistent with observation.  Type~Ia supernovae
supply about one half of the interstellar iron abundance in this model.

\section{Results}
\subsection{The solar neighbourhood}\label{results}

\begin{figure}[htb]
\begin{center}
\epsfig{file=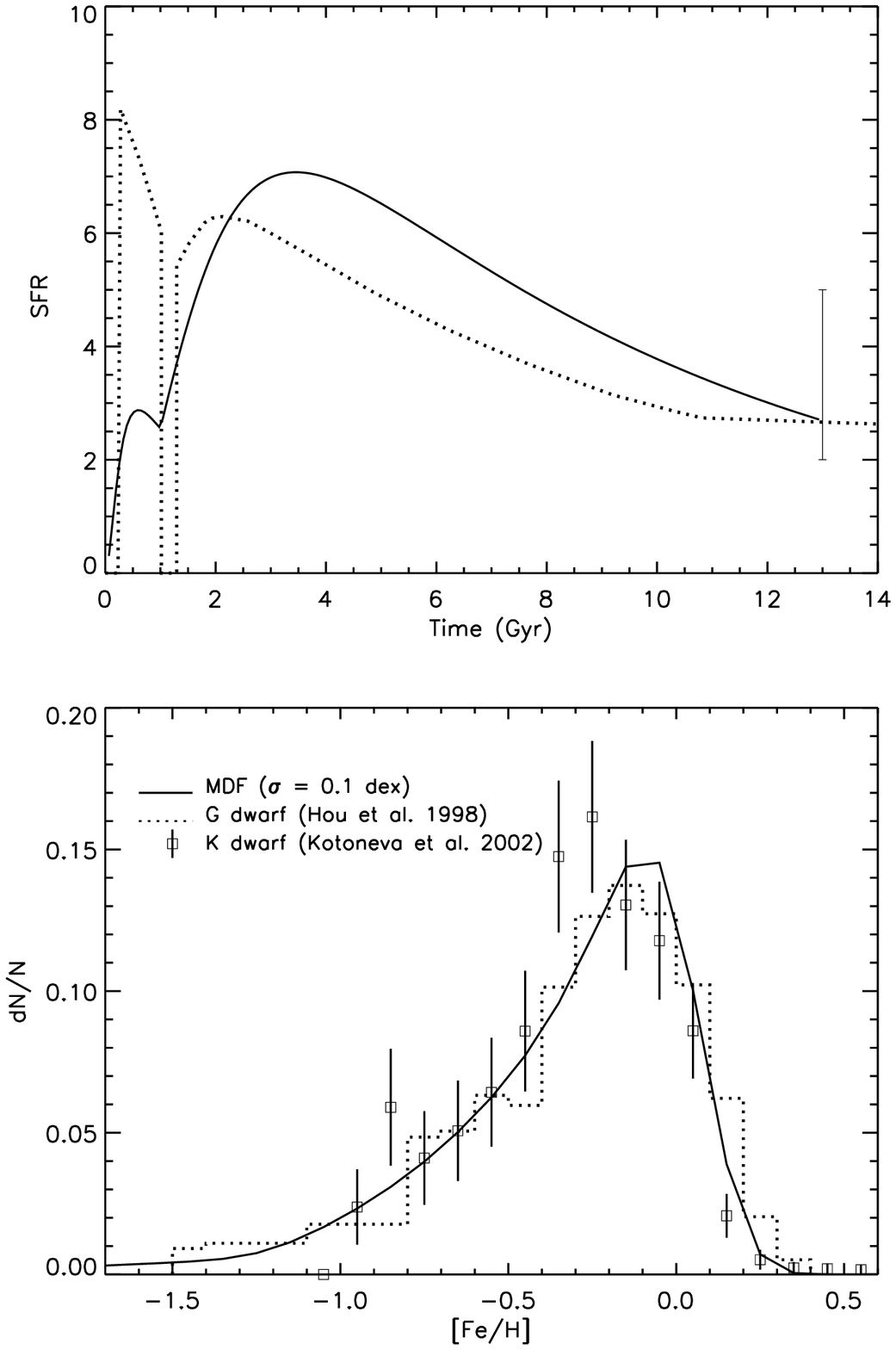,width=12cm}
\caption{\emph{Upper panel}: Evolution of the star formation rate (SFR)
  predicted by the dual-phase infall model described here (\emph{solid
    line}). The dotted line illustrates the stronger initial burst of
  star formation in the Chiappini et~al. (2001) dual-infall model.
  While the total integrated star formation differs between the two
  models by less than 20\%, during the first 1~Gyr the Chiappini
  et~al. model supplies $\sim$~2.5 times more stars than in our model.
  The SFR in the Chiappini et~al. disk phase also peaks 1.5~Gyr
  earlier than in our model.  The vertical bar denotes the observed
  limit on the present-day value.  \emph{Lower panel}: The calculated
  K-dwarf distribution (solid curve) is plotted against the datasets
  of Hou et~al. (1998 - \emph{dotted histogram}), and Kotoneva et~al.
  (2002 - \emph{open squares}).  The theoretical MDF has been
  convolved with a Gaussian of dispersion $\sigma$=0.1\,dex in [Fe/H],
  consistent with the known empirical uncertainties.}
\label{fig1} 
\end{center}
\end{figure}

The predicted star formation rate in the solar vicinity (\emph{solid
  line}) is shown as a function of time in the upper panel of
Figure~\ref{fig1} along with the range of measured values
(\emph{vertical bar}). The star formation rate in the halo peaks at
almost 3\,M$_{\odot}$~pc$^{-2}$~Gyr$^{-1}$ after 500~Myr. The
predicted present day rate of star formation in the thin disk of
2.7\,M$_{\odot}$~pc$^{-2}$~Gyr$^{-1}$ agrees with observations
(Chiappini et~al. 2001) and is about 2.5 times lower than the maximum
rate of star formation which occurred $\sim$\,10~Gyr ago, 2-3~Gyr after
the onset of the second infall phase.

The dotted line in the upper panel of Figure~\ref{fig1} depicts the
star formation rate given by the Chiappini et~al. (2001) dual-phase
Galactic chemical evolution model. Like the Milky Way model presented
in this paper, the Chiappini et~al. model is consistent with many
observational constraints, including the Kotoneva et~al. (2002)
empirical MDF. Despite both models satisfying key empirical
constraints such as the K-dwarf metallicity distribution, they do
predict very different properties for the halo, thick disk, and thin
disk components of the Milky Way. This is largely a consequence of
different star formation rates in the earliest epoch of Galactic
evolution. The predicted number of stars born during the first billion
years differs by a factor of 2.5 between the two models, whereas the
\emph{total} number of stars formed differs by less than 20\%. The
higher initial rate of star formation in Chiappini et~al. leads to a
more rapid enrichment of the interstellar medium, when compared
against the model presented here.  The implications of these
differences in terms of the Milky Way's stellar components are
quantified further below.

First, the model we have employed here is successful in reproducing
the new Kotoneva et~al. (2002) K-dwarf MDF, as shown in the lower
panel of Figure~\ref{fig1}. The predicted MDF has been convolved with
a $\sigma$\,=\,0.1\,dex Gaussian in [Fe/H], consistent with the
observational uncertainties in the data.  Plotted against the model
results (\emph{solid curve}) is the K-dwarf MDF from Kotoneva et~al.
(\emph{squares with error bars}) and the G-dwarf MDF from Hou et~al.
(1998) (\emph{dotted line histogram}). The model MDF provides a
satisfactory agreement with the Kotoneva et~al. dataset, although the
model peaks $\sim $~0.2~dex higher in [Fe/H].  The model also follows
closely the Hou et~al. G-dwarf MDF, which exhibits a longer metal-poor
tail than the K-dwarf distribution. While most authors measure the
peak of the local G-dwarf metallicity distribution to lie between
-0.2 and -0.3~dex (e.g. Wyse \& Gilmore 1995; Rocha-Pinto \& Maciel
1996), there is still not complete agreement.  Haywood (2001) found
that the MDF of a color-selected sample was centered on solar
metallicity, however Kotoneva et~al. showed that a bias toward
metal-rich stars might arise due to a systematic trend between
metallicity distribution and stellar color.

The complete absence of stars with [Fe/H]\,$<$\,$-$1 in the Kotoneva
et~al.\,(2002) sample is not reproduced by our model (\emph{solid
  curve}). In Chiappini et al. (1997; 2001) and Kotoneva et al.
(2002), any low-metallicity tail in the theoretical MDF is suppressed
by neglecting either: 1) stars with [Fe/H]\,$<$\,$-$1.2 (e.g.
Chiappini et al.  1997; Kotoneva et al. 2002); or 2) stars born before
t~=~1~Gyr (which effectively ignores stars with [Fe/H]\,$<$\,$-$0.6)
(e.g. Chiappini et al.  2001). In comparison, we have included {\it
  all} long-lived stars in the metallicity distribution function.
Stars with [Fe/H]\,$<$\,$-$1.2 represent a very small fraction of
present-day stars in the Chiappini et~al. model due to their strong
initial burst of star formation.  Efficient, early, star formation
leads to a metallicity [Fe/H]\,$=$\,$-$1.2 being reached on a very
short timescale ($<$\,0.1\,Gyr).  Such rapid initial enrichment is one
of the main differences between the present model and that of
Chiappini et~al.  We assume a more moderate initial phase of star
formation, leading to a metal-poor tail (i.e. [Fe/H]\,$<$\,$-$1.2) in
the MDF which contains $\sim$\,4\% of the stars.  The age associated
with this metal-poor tail is 12.5~Gyr, while the median age of our
predicted thin disk dwarfs is 7.5~Gyr. This is in excellent agreement
with Hansen et~al.  (2002) who derive ages of 7.3\,$\pm$\,1.5\,Gyr and
12.7\,$\pm$\,0.7\,Gyr for Galactic thin disk white dwarfs and for the
halo globular cluster M4, respectively.

\begin{figure}[htb]
\begin{center}
\epsfig{file=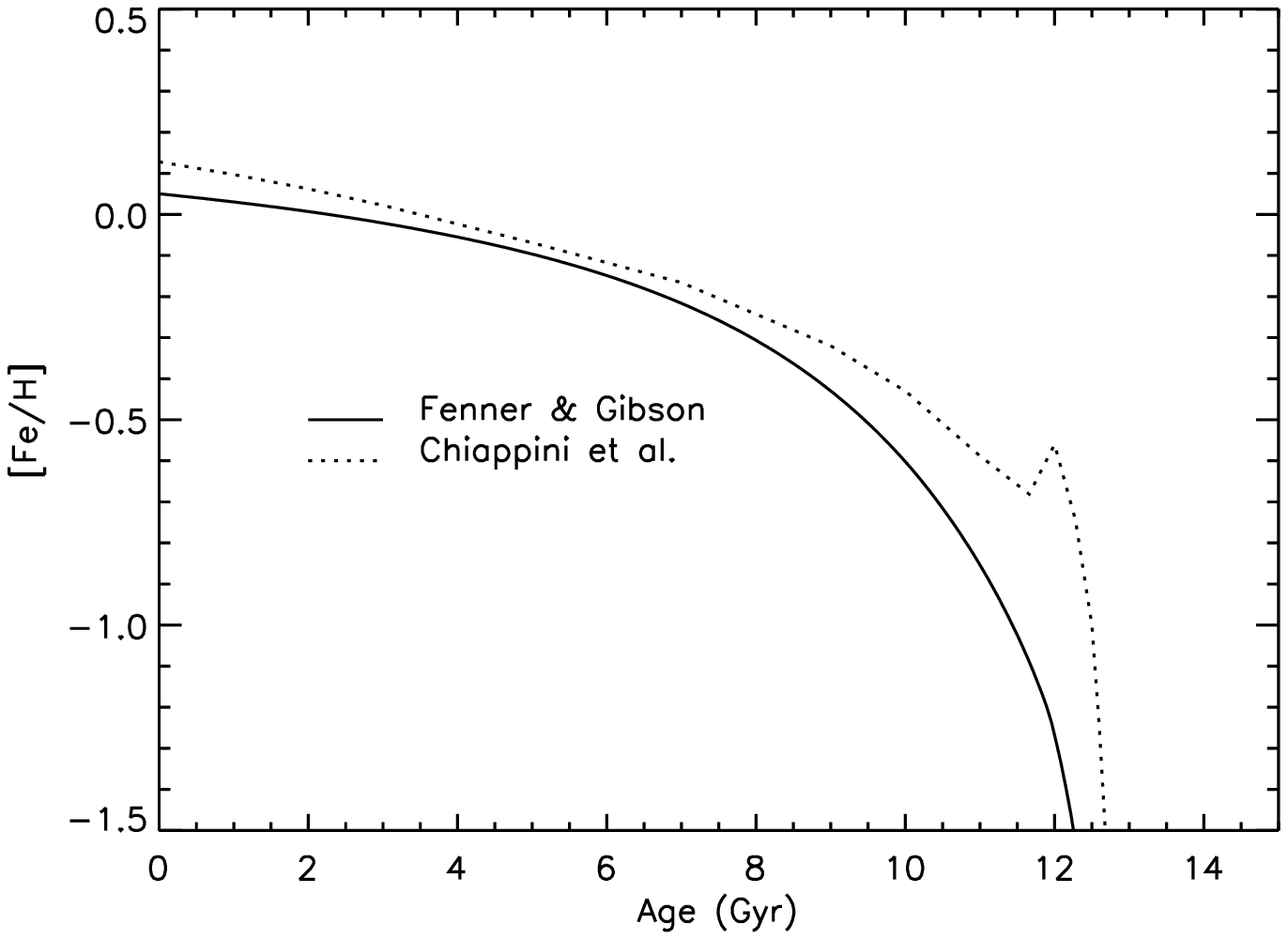}
\caption{The age-metallicity relation (AMR) in the solar neighbourhood 
predicted by our model (\emph{solid curve}) and by Chiappini et~al.
(2001 - \emph{dotted line}). While similar present-day iron
abundances are predicted by both models, in Chiappini et~al. the ISM
is polluted on a much shorter timescale. Our model assumes a
  Galactic age of 13~Gyr, whereas Chiappini et~al. (2001) take a value
of 14~Gyr. In comparing the two AMRs, we have
shifted the Chiappini et~al. curve by 1~Gyr such that both models span
13~Gyr.}
\label{fig2}
\end{center}     
\end{figure}

The age-metallicity relation (AMR) for our dual-infall model is shown in
Figure~\ref{fig2} (\emph{solid curve}). It takes $\sim$1\,Gyr to
enrich the interstellar medium in the solar neighborhood to a
metallicity of [Fe/H]\,=\,$-$1.25. In contrast, the Chiappini et~al.
(2001) model (\emph{dotted curve}) reaches the much greater
metallicity of [Fe/H]\,$\approx$\,$-$0.5 over the same time interval. The
AMR from Chiappini et~al. (2001) differs from that in the 1997 version
of their two-infall model, owing to different timescales for the first
episode of infall. Their halo/thick disk accumulates on a timescale of
2~Gyr in the 1997 model, with a 2~Gyr delay before the thin disk
starts to form. In the later model, these timescales are shortened to
$\sim$1\,Gyr, giving rise to an even steeper AMR than in the 1997
paper. The kink seen in the Chiappini et~al. AMR is due to the
metallicity of the ISM becoming diluted when the second phase of
infall commences at 1~Gyr. It is not until the thin disk gas surface
density reaches a threshold of 7\,M$_\odot$~pc$^{-2}$ that Chiappini
et~al. allow thin disk stars to form. No such star formation threshold
is incorporated in our model.

We wish to stress that caution must be employed when drawing
conclusions regarding the nature of {\it three} stellar population
components of the Milky Way (thin disk, thick disk, halo), when the
models are inherently limited by their {\it dual}-infall nature.  That
said, {\it if} one adopts a chemical criterion of
$-$1.0\,$<$\,[Fe/H]\,$<$\,$-$0.5 for the definition of ``thick disk'',
then the age of the thick disk predicted under the Chiappini et~al.
formalism would be centered upon 12.5\,Gyr, with a vanishingly small
age-spread ($<$\,0.2~Gyr), and possess kinematics closely aligned with
the first of the infall phases (i.e.  the halo).  Conversely, the
thick disk predicted under our formalism would be 2-3\,Gyr younger,
with an age-spread an order-of-magnitude larger ($\sim$\,2~Gyr), and
possess kinematics more closely resembling those of the second infall
phase (i.e. the disk).  The Kotoneva et~al.  (2002) sample of solar
neighborhood K-dwarfs identified only 2 of 431 stars as having halo
kinematics. Thus halo stars are rare, whereas the Chiappini model
would associate a large fraction of stellar mass with the halo
component.

In the Chiappini et al. model of Galaxy formation, the first infall
phase is associated with the formation of the halo and thick disk.
Thus these components are coeval and form prior to, and independently
of the thin disk. In such a scenario, the thick disk would have little
dispersion in age and the stars would exhibit primarily halo
kinematics. The model presented here associates the first infall phase
with the build up of the halo.  In contrast with Chiappini et al., the
formation of the thick disk overlaps with the early evolution of the
thin disk (during the second infall phase). Our thick disk, as defined
by metallicity, would have predominantly disk-like kinematics and a
broader range of ages.

While both formalisms clearly have problems in describing a
three-component system, we favour the scenario presented in this
paper, in which the intermediate metallicity population (thick disk)
has a typical age several Gyrs younger than the halo (as inferred by
observation - e.g. Chayboyer et al. 2000; Ibukiyama \& Arimoto 2002)
and kinematics akin to that of disk-like population, and not a
rapid-collapse halo population. In that sense, our model is more
consistent with the favored model of Galactic thick disk formation
whereby an existing stellar thin disk ``puffs up'' during the process
of a merger (e.g.  Wyse 2001).

\subsection{Lifetime effects and MDFs in external galaxies: A note of caution}\label{Harris}

Recent progress has been made in measuring the metallicity
distribution in the stellar halos of external galaxies (e.g. Durrell
et~al. 2001; Harris \& Harris 2000, 2002). The observed metallicity
distribution of these systems, which presumably formed early and on a
short timescale, is likely to be biased by the effect of finite
stellar lifetimes. In an attempt to quantify the discrepancy between
the observed and ``true'' MDF due to lifetime effects, we constructed
a ``toy'' model of the NGC~5128 halo using the chemical evolution code
described in this paper. This NGC~5128 halo model has some key
differences with respect to the Milky Way model described in this
paper. The main differences are: a single phase of star formation; a
shorter formation timescale ($\sim 1$~Gyr); and initial composition of
gas is primordial.

In stellar systems such as halos, a significant fraction of stars
classified as G-dwarfs may have evolved off the main-sequence. This
has the effect of preferentially removing older and therefore more
metal-poor stars from the observed present-day G-dwarf population. In
a similar way, the metal-poor tail in the measured G-dwarf MDF is
further suppressed by the shorter lifetimes of low mass, metal-poor
stars (e.g. Schaller et~al 1992).  On the other hand, the disk of our
own Galaxy has a more extended star formation history, with the
lengthier timescale and the higher proportion of young to old stars
acting to mitigate the ``loss'' of G-dwarfs from the present-day MDF.

\begin{figure}[hbt]
\epsfig{file=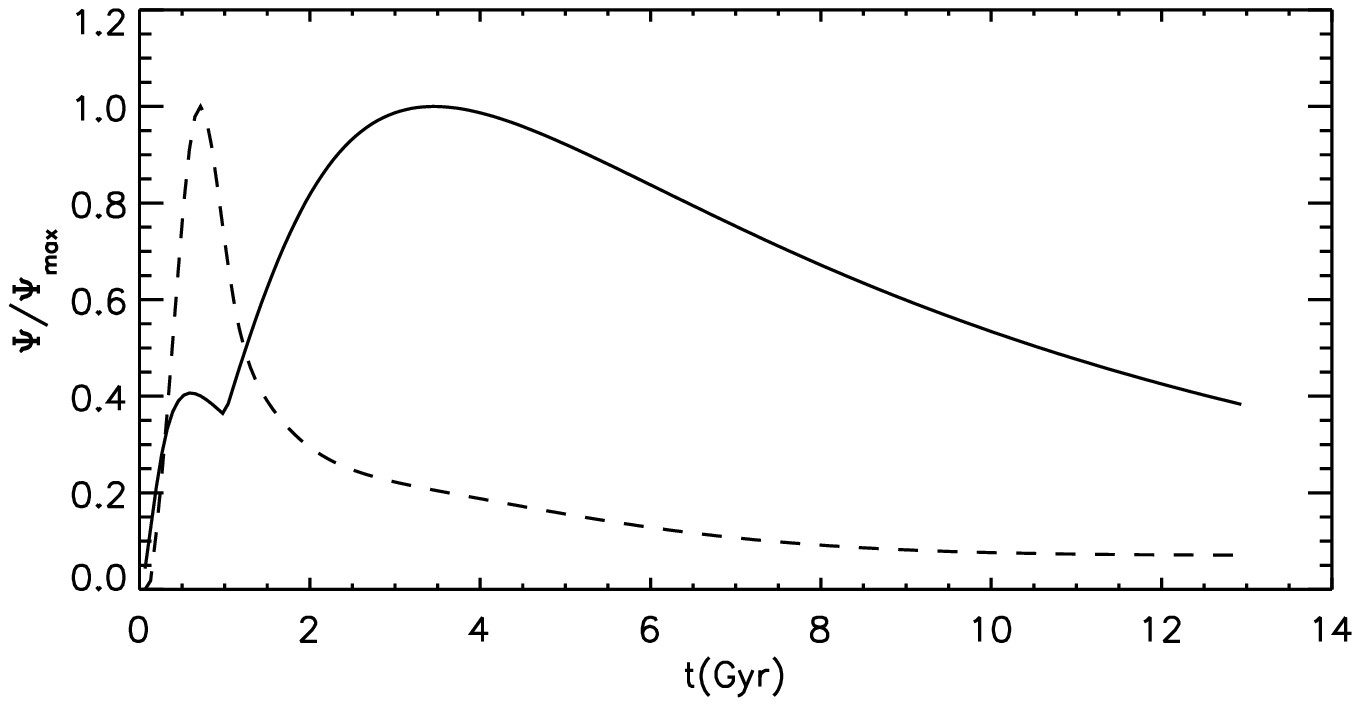}
\caption{Evolution of the star formation rate in a typical halo chemical
evolution model (\emph{dashed curve}) and in the Milky Way model
(\emph{solid curve}).  }
 \label{fig3}
\end{figure}

In Figure~\ref{fig3}, the star formation history of the Milky Way disk
is compared with a more burst-dominated star formation history, which
we here assume to be representative of ellipticals (or pure halo
systems).  The bulk of the stars in this burst model are born during
the first 1.5\,Gyr, after which the gas supply is depleted and the star
formation rate declines rapidly.  Conversely, it takes $>$ 5 Gyr for
the majority of stars to form in the disk model.

\begin{figure}[hbt]
\epsfig{file=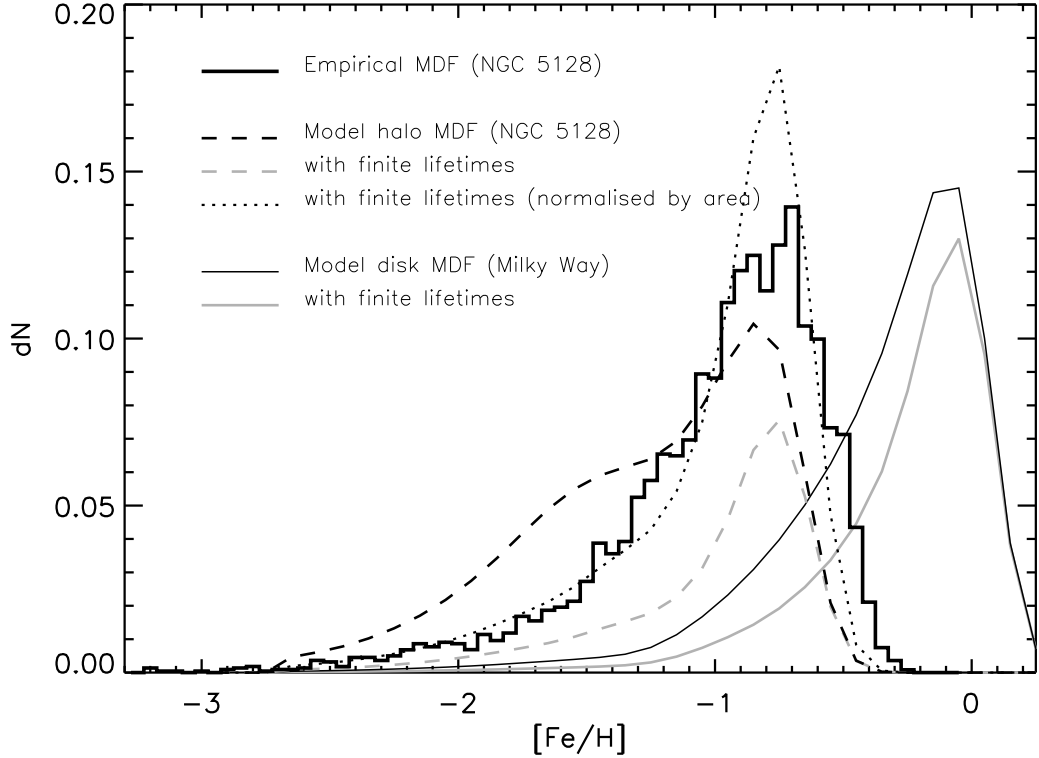}
\caption{Predicted metallicity distributions compared with
observations of the outer halo of NGC~5128 (Harris \& Harris 2000 -
\emph{histogram}).  We assumed the relation [Fe/H]=[m/H]$-$0.25 when
plotting the observed NGC~5128 MDF.  Dashed curves correspond to a
halo model, while solid curves represent the Milky Way model.  The
true MDF is indicated by black lines. Gray curves show the expected
MDF of the G-dwarfs that remain in the sample at $t$=15\,Gyr and
thus demonstrate the loss of stars from the sample due to finite
lifetimes. To illustrate the good fit, the dotted line presents the
finite lifetime halo model results, normalized to the NGC~5128 MDF
sample size.  Model predictions have been convolved with a Gaussian
of $\sigma$=0.1\,dex in metallicity.}
\label{fig4}
\end{figure}

Figure~\ref{fig4} shows the predicted MDF of truly long-lived stars
(\emph{black lines}) and G-dwarfs (defined as having masses in the
range 0.8\,$<$\,m/M$_{\odot}$\,$<$\,1.4) (\emph{gray lines}) for the halo
(\emph{dashed lines}) and Milky Way (\emph{solid lines}) models. The
observed MDF of the NGC~5128 outer halo (Harris \& Harris 2002) is
plotted as a histogram, where we have used the relation
[Fe/H]\,=\,[m/H]\,$-$\,0.25 to transform the published values of [m/H] into
units of [Fe/H]. The black lines show the MDF that is expected if all
G-dwarfs had lifetimes longer than the age of the galaxy (i.e., masses
m$<$0.8\,M$_\odot$).  These curves have been normalised to contain the
same area as the histogram. The gray curves show the MDF that would be
measured if the mass range of stars in the sample is
0.8\,$<$\,m/M$_{\odot}$\,$<$\,1.4.  These have been normalised to reflect the
loss of stars in each metallicity bin owing to evolution off the
main-sequence. The dotted line also shows the finite lifetime halo
model MDF, scaled to illustrate the fit to the data. The loss of
stars from a present-day sample of G-dwarfs can be seen to narrow the
MDF and shift the distribution toward higher metallicities.  While evident
in both models, this behaviour is more pronounced in environments such
as halos, where the bulk of the stars presumably formed $>$10\,Gyr
ago.

\begin{figure}[hbt]
\epsfig{file=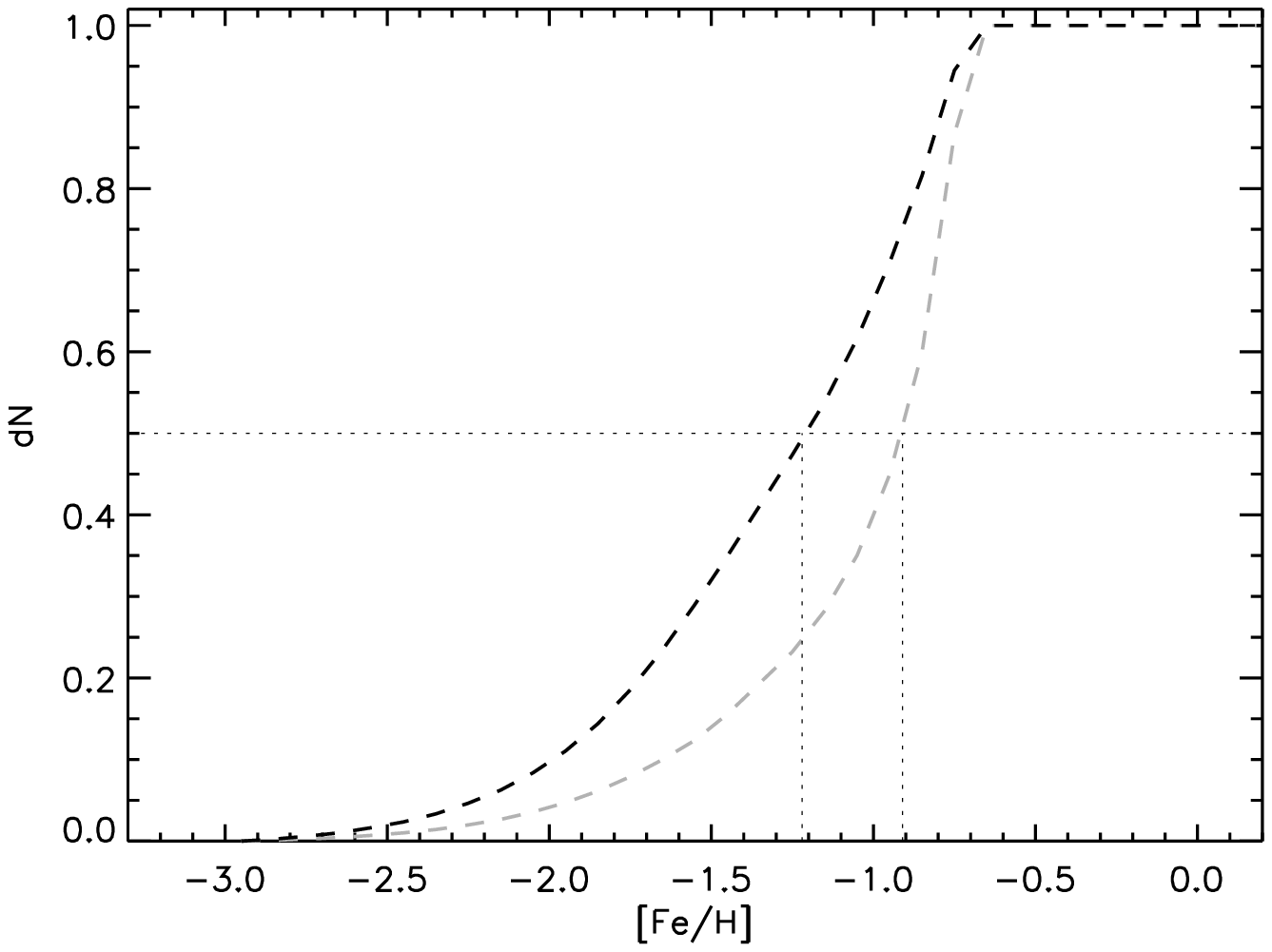}
\caption{Cumulative metallicity distribution function for a model
  NGC~5128 halo. The black dashed line corresponds to the metallicity
  distribution of long-lived stars, while the gray dashed line shows
  the expected observed MDF for a sample of stars in the mass range
  0.8$<$m/M$_{\odot}$$<$1.4, taking into account finite lifetimes. The
  median metallicity of a population corresponds to 0.5 on the
  vertical axis, as indicated by the horizontal dotted line. The
  evolution off the main sequence of older metal-poor G-dwarfs
  increases the median metallicity of the sample by $\sim 0.3$\,dex.}
\label{fig5}
\end{figure}

By ignoring lifetime effects, one can overestimate the true mean
metallicity of long-lived stars. To gauge the magnitude of this
effect, we plot the cumulative metallicity distribution in
Figure~\ref{fig5}. For this particular halo model, mass-dependent
stellar lifetimes serve to shift the median metallicity of the
population by $\Delta$[Fe/H]=$+$0.3.

\section{Conclusions}

We have reproduced the metallicity distribution of nearby K-dwarfs
using a model in which the Milky Way formed during two accretion
episodes. Several key differences between this and other dual-phase
infall models are (i) the assumption that the thin disk formed from
metal-enriched $\alpha$-enhanced extragalactic material, (ii) the
adoption of a new comprehensive set of stellar yields from Limongi
et~al. (2000; 2002) for calculating the contribution to the
enrichment of the ISM from massive stars, and (iii) a more gentle
initial phase of (halo) star formation, resulting in a shallower (but
still observationally consistent) age-metallicity relation.

Taking the stellar halo as having [Fe/H]\,$<$\,$-$1.2 we predict that this
component has an age of 12-13~Gyr and formed on a timescale of
$\sim$1\,Gyr. If stars with metallicity $-$1.0\,$<$\,[Fe/H]\,$<$\,$-$0.5 are
assigned to the thick disk, then the median age of this population is
predicted to be $\sim$10\,Gyr with a greater scatter in ages
($\sim$~$\pm$2\,Gyrs) than for the halo.  The age difference of $\sim
5$~Gyr between the halo and thin disk in our model is consistent with
empirical estimates based upon white dwarf luminosity functions
(Hansen et~al. 2002).

The model presented in this paper was compared with a similar
dual-phase infall model from Chiappini et~al. (2001) that also
reproduces the observed K-dwarf metallicity distribution. We showed
that even though differences in the rate of star formation in the
early Galaxy might not be apparent in present-day observables such as the
MDF, they do lead to different predictions for the ages and
kinematics of the halo, thick, and thin disk components. This degeneracy
is inherent in dual-infall models when applied to the modeling of 
three-component systems.

\section*{Acknowledgments}

It is with pleasure that we thank the referees Sean Ryan and Eira
Kotoneva for their very valuable comments on the manuscript.

We wish to thank Chris Flynn for many invaluable discussions.  We are
also grateful to Chris, Bill Harris, and Gretchen Harris, for
providing their observational metallicity distribution functions prior
to publication.

BKG acknowledges the financial support of the Australian Research
Council through its Large Research Grant Program (Grant ID
\#A00105171).

\section*{References}

  Argast,~D., Samland,~M., Thielemann,~F.~-K. \& Gerhard,~O.~E. 
2002, A\&A, 388, 842\\
  Bazan, G. \& Mathews, G.J. 1990, ApJ, 354, 644\\
  Carretta, E., Gratton, R. G. \& Sneden, C.  2000, A\&A, 356, 238\\
  Chaboyer B., Sarajedini, A. \& Armandroff, T.E. 2000, AJ, 120, 3102\\
  Chiappini, C., Matteucci, F. \& Gratton, R. 1997, ApJ, 477,
765\\
  Chiappini, C., Matteucci, F. \& Romano, D. 2001, ApJ, 554,
1044\\
  Durrell, P.R., Harris, W.E. \& Pritchet, C.J. 2001, AJ,
121, 2557\\
  ESA, 1997, The Hipparcos and Tycho Catalogues, ESA, SP-1200\\
  Fenner, Y. \& Gibson, B.K. 2003, in preparation\\
  Fenner, Y., Gibson, B.K. \& Limongi, M. 2002, Ap\&SS, 281, 537\\
  Gibson, B.K., Giroux, M.L., Penton, S.V., Stocke, J.T.,
Shull, J.M. \& Tumlinson, J. 2001, AJ, 547, 3280\\
  Gilmore, G., Wyse, R.F.G. \& Jones, J.B. 1995, AJ, 109, 1095\\
  Goswami, A. \& Prantzos, N. 2000, A\&A, 359, 191\\
 Hansen, B.M.S., Brewer, J., Fahlman, G.G., Gibson, B.K.,
 Ibata, R., Limongi, M., Rich, R.M., Richer, H.B., Shara, M.M. \& Stetson, P.B.
 2002, ApJ, 574, L155\\
  Ibukiyama, A. \& Arimoto, N. 2002, A\&A, 394, 927\\
  Harris, G.L.H. \& Harris, W.E. 2000, AJ, 120, 2423\\
  Harris, W.E. \& Harris, G.L.H. 2002, AJ, 123, 3108\\
  Haywood, M. 2001, MNRAS, 325, 1365\\
  Hou, J.L., Chang, R. \& Fu, C. 1998, in Pacific Rim
 Conference on Stellar Astrophysics, ed. K.L. Chan, K.S. Cheng \&
 H.P. Singh (San Francisco: ASP), p.~143\\
  Kotoneva, E., Flynn, C., Chiappini, C. \& Matteucci,
 F. 2002, MNRAS, 336, 879\\
  Kroupa, P., Tout, C.A. \& Gilmore, G. 1993, MNRAS, 262,
545\\
  Limongi, M., Straniero, O. \& Chieffi, A., 2000, ApJS, 129, 625\\
  Limongi, M. \& Chieffi, A., 2002 PASA, 19, 246\\
  Melendez, J., Barbuy, B. \& Spite, F. 2001, ApJ, 556, 858\\
  McWilliam, A. 1997, ARAA, 35, 503 \\
  Norris, J.E. \& Ryan, S.G., 1991, ApJ, 380, 403\\
  Prantzos, N. \& Silk, J. 1998, ApJ, 507, 229\\
  Renzini, A. \& Voli, M. 1981, A\&A, 94, 175\\
  Rocha-Pinto, H.-J. \& Maciel, W.J. 1996, MNRAS, 279, 447\\
  Rocha-Pinto, H.-J. \& Maciel, W.J. 1997, A\&A, 325, 523\\
  Romano, D., Matteucci, F., Salucci, P. \& Chiappini, C. 2000, ApJ,
  539, 235\\
  Ryan, S. G. Norris, J. E. \& Beers, T. C. 1996, ApJ, 471,
  254\\
  Salpeter, E.E. 1955, ApJ, 121, 161\\
  Scalo, J.M. 1986, Fund. Cosm. Phys., 11, 1\\
  Schaller, G., Schaerer, D., Meynet, G. \& Maeder,
A. 1992, A\&AS, 96, 269\\
  Sembach, K.R., Gibson, B.K., Fenner, Y. \& Putman, M.E.
2002, ApJ, 572, 178\\
  Thielemann, F.-K., Nomoto, K., Hashimoto, M., 1993, in Prantzos, N.,
  Vangiono-Flam, E., Casse, M., eds, Origin and Evolution of the
  Elements, Cambridge Univ. Press, Cambridge, p. 297\\
  Tinsley, B.M. 1980, Fund. Cosm. Phys., 5, 287\\
  Wakker, B.P., Howk, J.C., Savage, B.D., van~Woerden, H.,
Tufte, S.L., Schwarz, U.J., Benjamin, R., Reynolds, R. J., Peletier, R. F. \&
Kalberla, P. M. W. 1999, Nature, 402, 388\\
  Woosley, S. E. \& Weaver, T. A. 1995, ApJS, 101, 181\\
  Wyse, R.F.G. 2001, in ASP Conf. Ser. 230, Galaxy Disks and Disk Galaxies, eds. J. G. Funes \& E. M. Corsini (San Francisco: ASP), 71\\
  Wyse, R.F.G. \& Gilmore, G. 1995, AJ, 110, 2771\\

\end{document}